# Near-field Analysis of Superluminally Propagating Electromagnetic and Gravitational Fields

**William D. Walker**
Örebro University, Department of Technology, Sweden
Research papers [1]
william.walker@tech.oru.se

## Abstract

A near-field analysis based on Maxwell's equations is presented which indicates that the fields generated by both an electric and a magnetic dipole or quadrapole, and also the gravitational waves generated by a quadrapole mass source propagate superluminally in the nearfield of the source and reduce to the speed of light as the waves propagate into the farfield. Both the phase speed and the group speed are shown to be superluminal in the nearfield of these systems. Although the information speed is shown to differ from group speed in the nearfield of these systems, provided the noise of the signal is small and the modulation method is known, the information can be extracted in a time period much smaller than the wave propagation time, thereby making the information speed only slightly less than the superluminal group speed. It is shown that relativity theory indicates that these superluminal signals can be reflected off of a moving frame causing the information to arrive before the signal was transmitted (i.e. backward in time). It is unknown if these signals can be used to change the past.

## Introduction

The electromagnetic fields generated by an oscillating electric dipole have been theoretically studied by many researchers using Maxwell's equations and are known to yield the following well-known results:

## System differential equation

**Fig. 1**

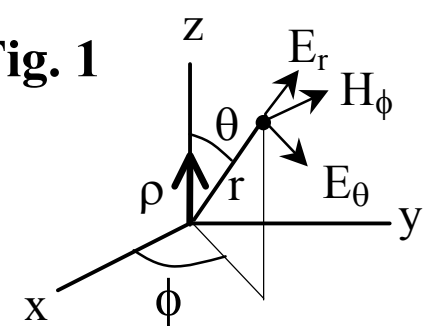

System PDE

$$\nabla^2 V - \frac{1}{c^2}\frac{\partial^2 V}{\partial t^2} = \frac{-\rho}{\varepsilon_o} \qquad (1)$$

Variable definitions
$E_r$ = Radial electric field
$E_\theta$ = Transverse electric field
$H_\phi$ = Transverse magnetic field
$V$ = Scalar potential
$\rho$ = Charge density
$\varepsilon_o$ = Free-space permitivity
$\nabla^2$ = Laplacian
$c$ = Speed of light

## Field analysis

Solving the homogeneous equation (Eq. 1) for a dipole source yields [2, 3, 4, 5, 6]:

$$V = N\,Cos(\theta)\left[\frac{1}{kr} + \frac{i}{(kr)^2}\right]e^{ikr-\omega t} \qquad (2)$$

The fields can then be calculated using the following relations [6]:

$$B = r \times \nabla V \qquad E = \frac{ic}{\omega}(\nabla \times B) \qquad (3)$$

yielding: $$E_r = \frac{\rho Cos(\theta)}{2\pi\varepsilon_o r^3}\left[1 - i(kr)\right]e^{i(kr-\omega t)}$$

$$E_\theta = \frac{\rho Sin(\theta)}{4\pi\varepsilon_o r^3}\left[\left\{1-(kr)^2\right\} - i(kr)\right]e^{i(kr-\omega t)} \qquad H_\varphi = \frac{\omega\rho Sin(\theta)}{4\pi\, r^2}\left[-kr - i\right]e^{i(kr-\omega t)} \qquad (4)$$

## Phase analysis

The general form of the electromagnetic fields generated by a dipole is:

$$Field \propto (x+iy)\cdot e^{i[kr-\omega t]}$$

If the source is modelled as $Cos(\omega t)$, the resultant generated field is:

$$Field \propto Mag \cdot Cos[\{kr+ph\} - \omega t] = Mag \cdot Cos(\theta - \omega t)$$

where: $\mathrm{Mag} = \sqrt{\mathrm{x}^2 + y^2}$

It should be noted that the formula describing the phase is dependant on the quadrant of the complex vector.

$$\theta_1 = kr + Tan^{-1}\left(\frac{y}{x}\right) \qquad \theta_2 = kr - Cos^{-1}\left(\frac{x}{\sqrt{x^2+y^2}}\right) \tag{5}$$

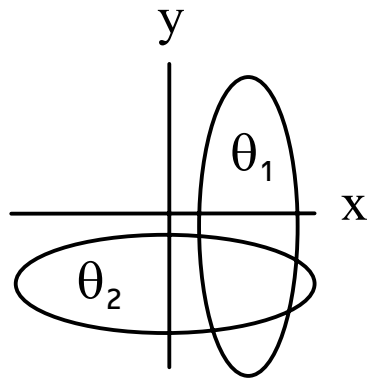

## Phase speed analysis

Phase speed can be defined as the speed at which a wave composed of one frequency propagates. The phase speed ($c_{ph}$) of an oscillating field of the form $Sin(\omega t - kr)$, in which $k = k(\omega, r)$, can be determined by setting the phase part of the field to zero, differentiating the resultant equation, and solving for $\partial r / \partial t$.

$$\frac{\partial}{\partial t}(\omega t - kr) = 0 \qquad \therefore \omega - k\frac{\partial r}{\partial t} - r\frac{\partial k}{\partial r}\frac{\partial r}{\partial t} = 0 \qquad \therefore c_{ph} = \frac{\partial r}{\partial t} = \frac{\omega}{k + r\dfrac{\partial k}{\partial r}} \tag{6}$$

Differentiating $\theta \equiv -kr$ with respect to r yields: $\dfrac{\partial \theta}{\partial r} = -k - r\dfrac{\partial k}{\partial r}$ (7)

Combining these results and inserting the far-field wave number ($k = \omega/c_o$) yields:

$$c_{ph} = -\omega \Big/ \frac{\partial \theta}{\partial r} = \boxed{-c_o k \Big/ \frac{\partial \theta}{\partial r}} \tag{8}$$

## Group speed analysis

The group speed of an oscillating field of the form $Sin(\omega t - kr)$, in which $k = k(\omega, r)$, can be calculated by considering two Fourier components of a wave group [7]: $Sin(\omega_1 t - k_1 r) + Sin(\omega_2 t - k_2 r) = Sin(\Delta\omega t - \Delta k r)\ Sin(\omega t - kr)$ (9) in which: $\Delta\omega = \dfrac{\omega_1 - \omega_2}{2}$, $\Delta k = \dfrac{k_1 - k_2}{2}$, $\omega = \dfrac{\omega_1 + \omega_2}{2}$, $k = \dfrac{k_1 + k_2}{2}$

The group speed ($c_g$) can then be determined by setting the phase part of the modulation component of the field to zero, differentiating the resultant equation, and solving for $\partial r / \partial t$:

$$\frac{\partial}{\partial t}(\Delta\omega t - \Delta k r) = 0 \qquad \therefore \Delta\omega - \Delta k\frac{\partial r}{\partial t} - r\frac{\partial \Delta k}{\partial r}\frac{\partial r}{\partial t} = 0 \qquad \therefore c_g = \frac{\partial r}{\partial t} = \frac{\Delta\omega}{\Delta k + r\dfrac{\partial \Delta k}{\partial r}} \tag{10}$$

Differentiating $\Delta\theta \equiv -\Delta k r$ with respect to r yields: $\dfrac{\partial \Delta\theta}{\partial r} = -\Delta k - r\dfrac{\partial \Delta k}{\partial r}$ (11)

Combining these results and using the far-field wave number ($k = \omega/c_o$) yields:

$$c_g = -\Delta\omega \Big/ \frac{\partial \Delta\theta}{\partial r} = -\left[\frac{\partial}{\partial r}\frac{\Delta\theta}{\Delta\omega}\right]^{-1} \qquad \therefore \lim_{\frac{\Delta\theta}{\Delta\omega}\ small} c_g = -\left[\frac{\partial^2\theta}{\partial r \partial\omega}\right]^{-1} = \boxed{-\frac{1}{c_o}\left[\frac{\partial^2\theta}{\partial r \partial k}\right]^{-1}} \tag{12}$$

It should be noted that other derivations of the above phase and group speed relations are available in previous publications by the author [8, 9, 10, 11, 12] and in the following well-known reference [13].

## Wave propagation analysis of near-field electric dipole EM fields

To determine how the EM fields propagate in an electric dipole system, one can apply the above phase and group speed relations (Eq. 8, 12) to the known theoretical solution of an electric dipole (Eq. 4).

## Radial electric field (Er) solution

$$y = -kr \qquad x = 1$$

$$\theta = kr - Tan^{-1}(kr) \underset{kr<<1}{\approx} -\frac{1}{3}(kr)^3 \tag{13}$$

$$c_{ph} = c_o\left(1 + \frac{1}{(kr)^2}\right) \underset{kr<<1}{\approx} \frac{c_o}{(kr)^2} \underset{kr>>1}{\approx} c_o \tag{14}$$

$$c_g = \frac{c_o\left(1+(kr)^2\right)^2}{3(kr)^2 + (kr)^4} \underset{kr<<1}{\approx} \frac{c_{ph}}{3} \underset{kr>>1}{\approx} c_o \tag{15}$$

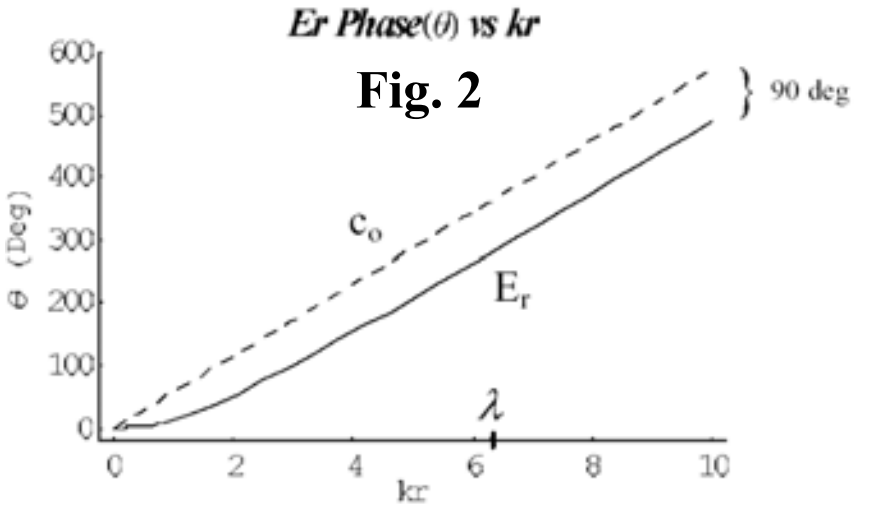

Fig. 2

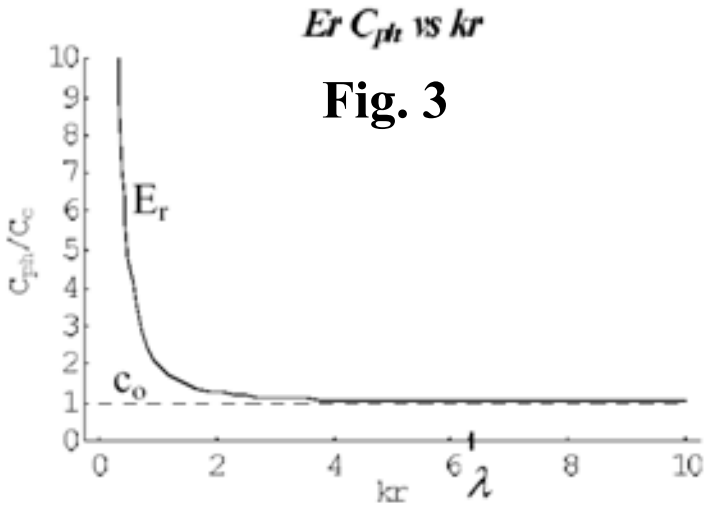

Fig. 3

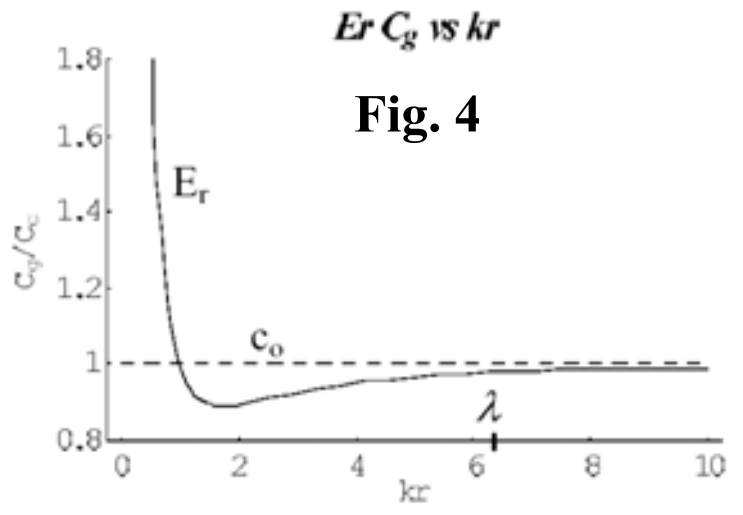

Fig. 4

## Transverse electric field ($E_\theta$) solution

$$y = -kr \qquad x = 1-(kr)^2$$

$$\theta = kr - Cos^{-1}\left(\frac{1-(kr)^2}{\sqrt{1-(kr)^2+(kr)^4}}\right) \tag{16}$$

$$c_{ph} = c_o\left(\frac{1-(kr)^2+(kr)^4}{-2(kr)^2+(kr)^4}\right) \tag{17}$$

$$c_g = \frac{c_o\left(1-(kr)^2+(kr)^4\right)^2}{-6(kr)^2+7(kr)^4-(kr)^6+(kr)^8} \tag{18}$$

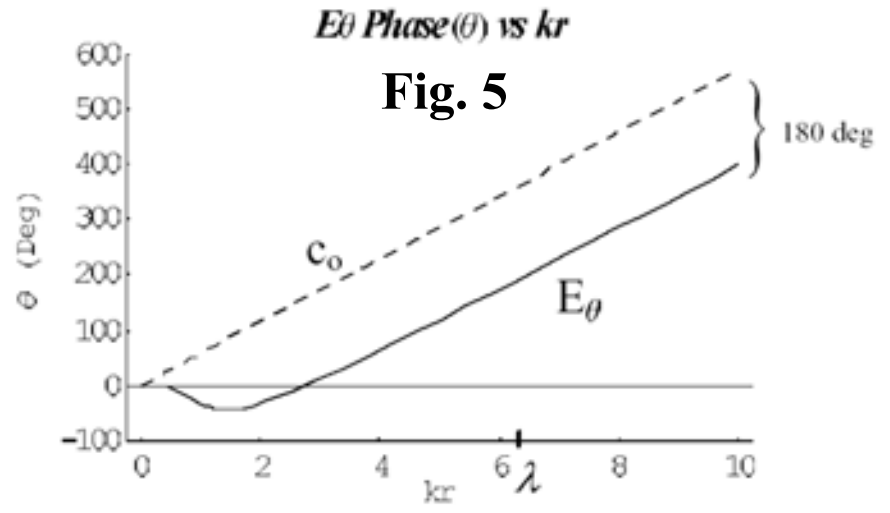

Fig. 5

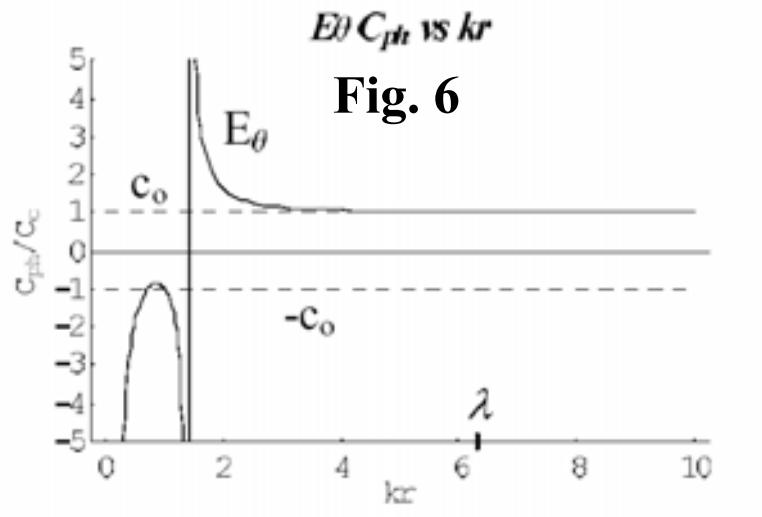

Fig. 6

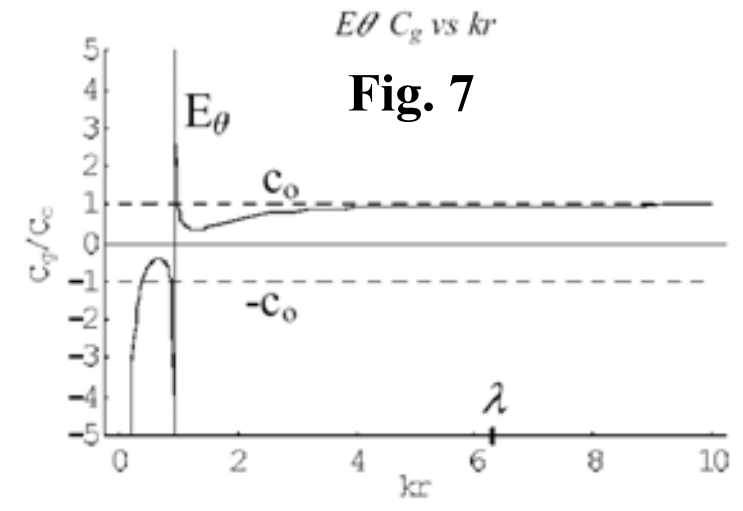

Fig. 7

## Transverse magnetic field ($H_\phi$) solution

$$y = -1 \qquad x = -kr$$

$$\theta = kr - Cos^{-1}\left(\frac{-kr}{\sqrt{1+(kr)^2}}\right) \tag{19}$$

$$c_{ph} = c_o\left(1 + \frac{1}{(kr)^2}\right) \tag{20}$$

$$c_g = \frac{c_o\left(1+(kr)^2\right)^2}{3(kr)^2+(kr)^4} \tag{21}$$

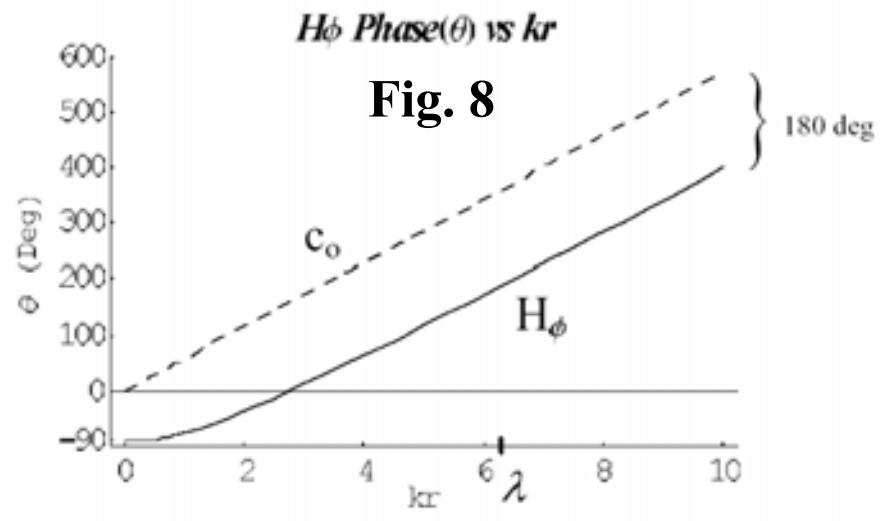

Fig. 8

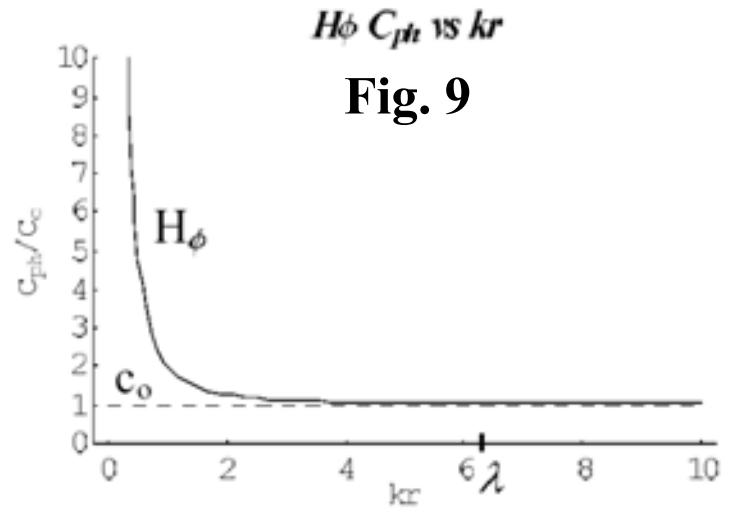

Fig. 9

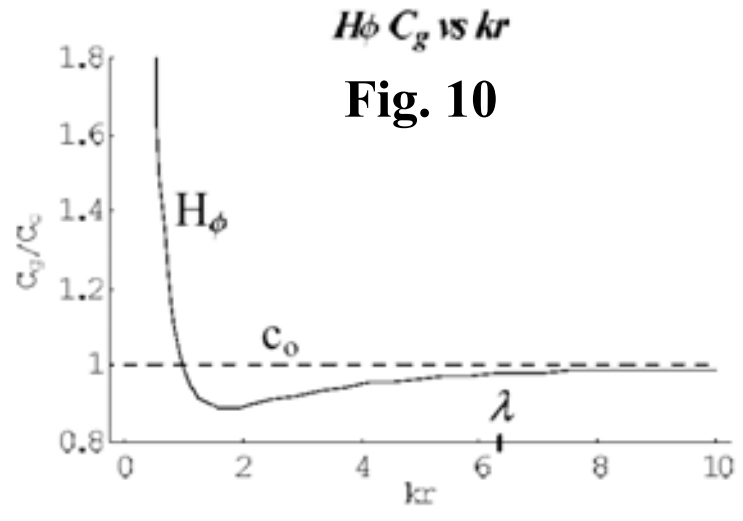

Fig. 10

## Experimental verification of $E_\theta$ solution

A simple experimental setup using two dipole antennas, a 437 MHz (68.65 cm wavelength), 2 watt sinusoidal transmitter, and a 500MHz digital oscilloscope has been developed to verify qualitatively the transverse electric field phase vs. distance plot predicted from standard EM theory (Eq. 16, Fig. 5). The phase shift of the received antenna signal (Rx) relative to the transmitted signal (Tx) was measured with the oscilloscope as the distance between the antennas ($r_{el}$) was changed from 5 cm to 70 cm in increments of 5 cm. The data was then curve fit with a 3rd order polynomial. The phase speed and group speed vs. distance curves were then generated by differentiating the resultant curve fit equation with respect to space and using (Eq. 8, 12). Refer to a previous paper written by the author for more details [8].

Experimental setup

**Fig. 11**

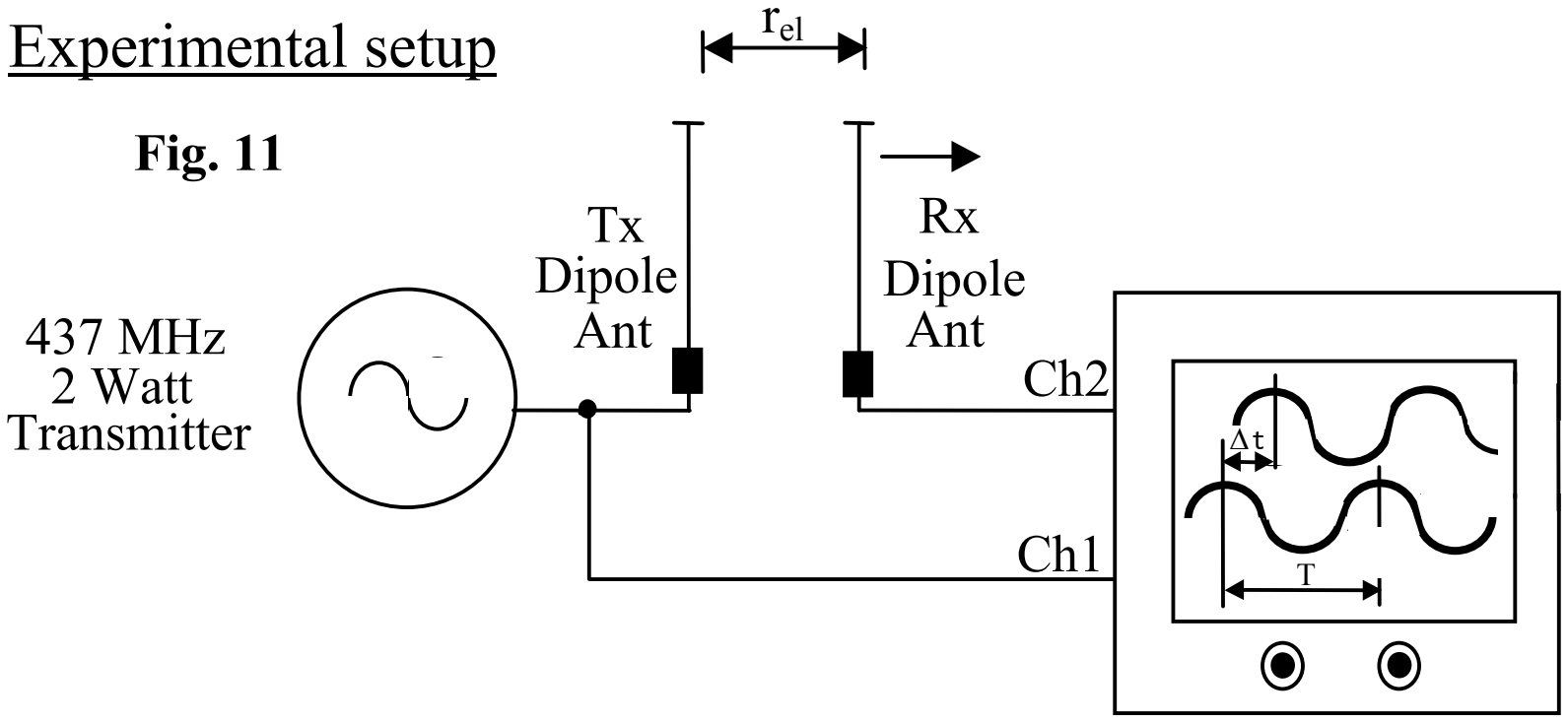

Experimental data

**Fig. 12**

| Data # | $r_{el}$ (cm) | Ph (Deg) |
|---|---|---|
| 0 | 0 | 140.0 |
| 1 | 5 | 111.7 |
| 2 | 10 | 102.4 |
| 3 | 15 | 108.6 |
| 4 | 20 | 121.0 |
| 5 | 25 | 136.6 |
| 6 | 30 | 155.2 |
| 7 | 35 | 170.7 |
| 8 | 40 | 195.5 |
| 9 | 45 | 211.0 |
| 10 | 50 | 245.2 |
| 11 | 55 | 282.4 |
| 12 | 60 | 316.6 |
| 13 | 65 | 325.9 |
| 14 | 70 | 366.2 |

$E_\theta$ phase plot similar to Fig. 5

**Fig. 13**

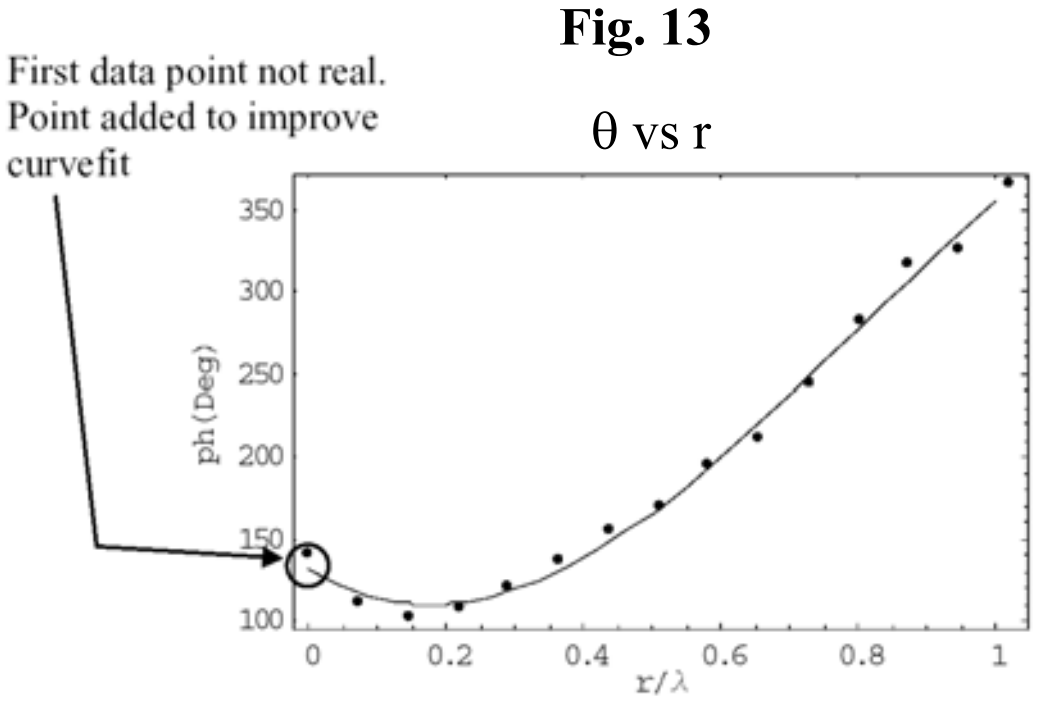

$E_\theta$ curvefit equation

$$ph = (132.2) + (-262.5)r_{el} + (838.9)r_{el}^2 + (-353.4)r_{el}^3 \quad (22)$$

Resultant $E_\theta$ phase speed and group speed plots - very similar to the predicted plots (Fig. 6, 7)

$c_{ph}/c_o$ vs $r/\lambda$

**Fig. 14**

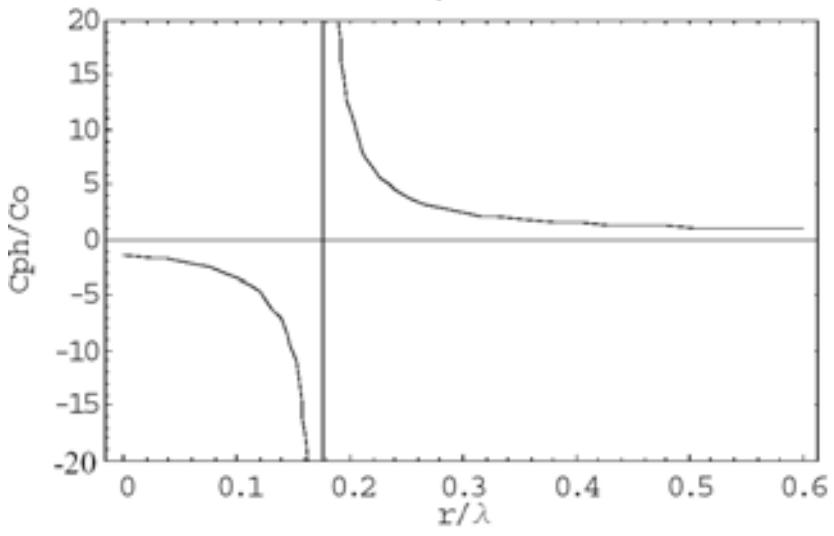

$c_g/c_o$ vs $r/\lambda$

**Fig. 15**

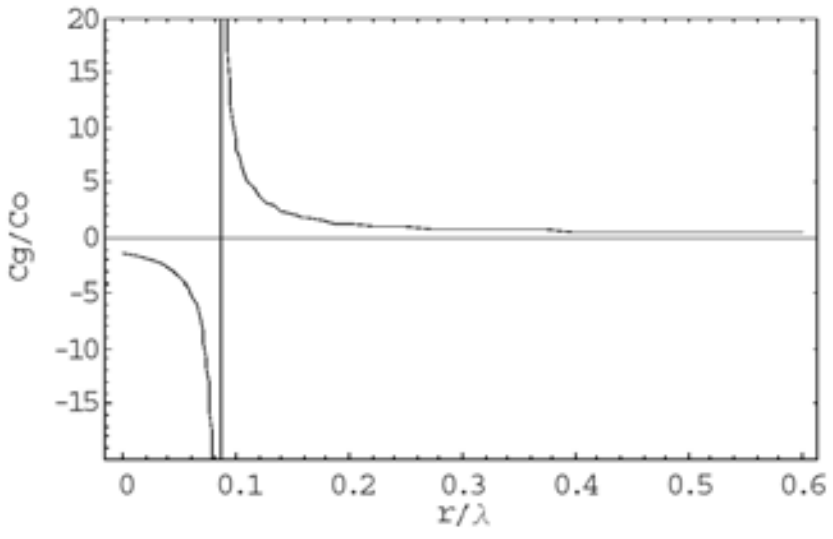

## Other wave propagation systems with similar superluminal behavior

### Magnetic dipole

Theoretical analysis of a magnetic dipole reveals that the system is governed by the same partial differential equation as the electric dipole with the E and B fields interchanged [2, 4]. The resulting fields are found to be the same as the fields generated by an electric dipole (Eq. 4) and therefore the phase speed and group speed of these fields are the same as (Eq. 13 - 21), except that the E and B fields are interchanged.

### Electric and magnetic quadrapole

Using the same method of analysis as was done for the electric dipole, the scalar potential for an electric quadrapole is found to be [3]:

$$V = -N\left(i\left(\frac{1}{kr} - \frac{3}{(kr)^3}\right) - \frac{3}{(kr)^2}\right)\left(3Cos^2\theta - 1\right)e^{i(kr-\omega t)} \tag{23}$$

The fields can then be calculated using the following relations [6]:

$$B = r \times \nabla V \qquad E = \frac{ic}{\omega}(\nabla \times B) \tag{24}$$

yielding:

$$B_\phi = N\left[\left(\frac{1}{kr} - \frac{3}{(kr)^3}\right)i - \frac{3}{(kr)^2}\right]\left[Cos(\theta)Sin(\theta)\right]\, e^{i(kr-\omega t)} \tag{25}$$

$$E_r = N\left[\left(\frac{-3}{(kr)^3}\right)i - \frac{1}{(kr)^2} + \frac{3}{(kr)^4}\right]\left[3Cos^2(\theta) - 1\right]\, e^{i(kr-\omega t)} \tag{26}$$

$$E_\theta = 6N\left[\left(\left(\frac{1}{kr}\right) - \frac{6}{(kr)^3}\right)i + \frac{6}{(kr)^4} - \frac{3}{(kr)^2}\right]\left[Cos(\theta)Sin(\theta)\right]e^{i(kr-\omega t)} \tag{27}$$

where: $N = \dfrac{Qs^2k^4}{4\pi\varepsilon_o}$ in all the above solutions.

s = Dipole length, Q= Charge
k = Wave number, $\varepsilon_o$ = Permitivity

The fields for a magnetic quadrapole are also the same, with the E and B fields interchanged. The phase and group speed of these fields can be determined using the relations (Eq. 8, 12). These results will be presented in the next section (Eq. 34 - 42).

### Gravitational quadrapole

For weak and slowly varying gravitational fields, Einstein's equation becomes linearized and reduces to [14, 15]:

$$\nabla^2 V - \frac{1}{c^2}\frac{\partial^2 V}{\partial t^2} = 4\pi G\rho \tag{28}$$

Where: $\rho$ = Mass density, V = Gravitational potential, G = Gravitational constant, $\nabla^2$ = Laplacian, c = Speed of light, t = Time

Except for the source term, the partial differential equation of the potential is the same as that of an oscillating charge (Eq. 1). Because of this similarity one can then use the oscillating charge solutions by simply substituting: $\varepsilon_o = -1/(4\pi G)$. In addition, because momentum is conserved, a moving mass must push off another mass. The gravitational field generated by the secondary mass adds to the gravitational fields generated by the moving mass, resulting in a linear quadrapole source. The gravitational fields generated by an oscillating mass are therefore of the same form as the fields generated by an electric quadrapole (Eq. 23 - 27).

$$V = -N\left(i\left(\frac{1}{kr} - \frac{3}{(kr)^3}\right) - \frac{3}{(kr)^2}\right)\left(3Cos^2\theta - 1\right)e^{i(kr-\omega t)} \qquad (29)$$

It is known that for weak and slowly varying gravitational fields, General Relativity theory reduces to a form of Maxwell's equations [14]. The fields can then be calculated using the following relations:

$$B = r \times \nabla V \qquad\qquad E = \frac{ic}{\omega}(\nabla \times B) \qquad (30)$$

where (E) is the gravitational force vector and (B) is the solenoidal gravitational force vector. The constant (N) can be determined by substituting the relation: $\varepsilon_o = -1/(4\pi G)$ into the value of (N) used in the electric quadrapole (Eq. 27). In addition, this result can be checked by looking at the static quadrapole solution and comparing it to the above solutions in the limit ($kr \to 0$). The results yield:

$$B_\phi = N\left[\left(\frac{1}{kr} - \frac{3}{(kr)^3}\right)i - \frac{3}{(kr)^2}\right]\left[Cos(\theta)Sin(\theta)\right]e^{i(kr-\omega t)} \qquad (31)$$

$$E_r = N\left[\left(\frac{-3}{(kr)^3}\right)i - \frac{1}{(kr)^2} + \frac{3}{(kr)^4}\right]\left[3Cos^2(\theta) - 1\right]e^{i(kr-\omega t)} \qquad (32)$$

$$E_\theta = 6N\left[\left(\left(\frac{1}{kr}\right) - \frac{6}{(kr)^3}\right)i + \frac{6}{(kr)^4} - \frac{3}{(kr)^2}\right]\left[Cos(\theta)Sin(\theta)\right]e^{i(kr-\omega t)} \qquad (33)$$

where: $N = -G\rho\, s\, k^4$, G = Grav const., s = Dipole length, k = Wave number

The phase and group speed relations for these fields can then be determined by using the phase and group speed equations derived earlier in the paper (Eq. 8, 12). It should be noted that all the plots look very similar to those of an electric dipole (ref. p.3).

**$B_\phi$ phase, phase speed, group speed analysis**

$$ph = kr - ArcTan\left[\frac{kr}{3} - \frac{3}{kr}\right] \underset{kr \ll 1}{\approx} \frac{\pi}{2} + \frac{1}{45}(kr)^5 + O(kr)^7 \qquad (34)$$

$$Cph = c_o\left(1 + \frac{3}{(kr)^2} + \frac{9}{(kr)^4}\right) \qquad (35)$$

$$Cg = \frac{c_o\left[3(kr)^2 + (kr)^4 + 9\right]^2}{(kr)^4\left[45 + 9(kr)^2 + (kr)^4\right]} \qquad (36)$$

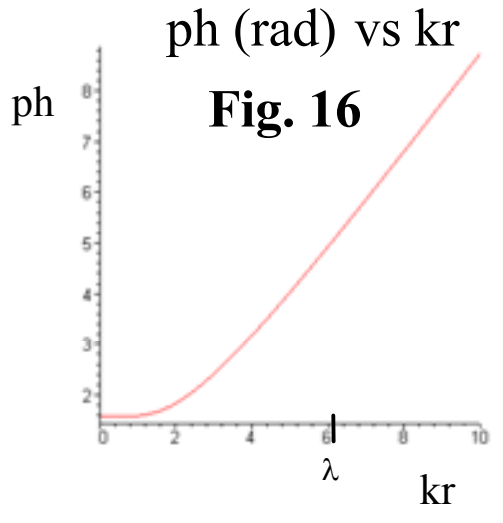

**Fig. 16**

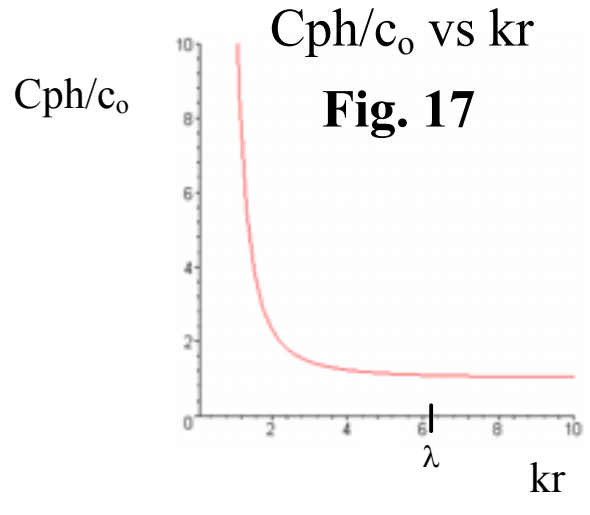

**Fig. 17**

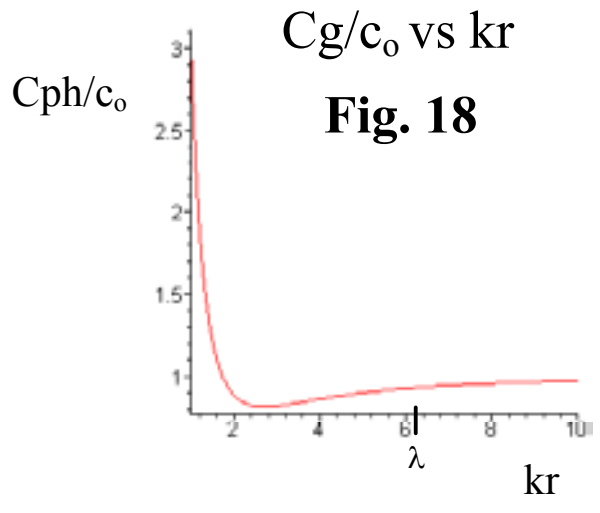

**Fig. 18**

## Er phase, phase speed, group speed analysis

$$ph = kr + ArcTan\left[\frac{3}{kr - \frac{3}{kr}}\right] \underset{kr \ll 1}{\approx} \frac{1}{45}(kr)^5 + O(kr)^7 \tag{37}$$

$$Cph = c_o\left(1 + \frac{3}{(kr)^2} + \frac{9}{(kr)^4}\right) \tag{38}$$

$$Cg = \frac{c_o\left[(\mathrm{kr})^4 + 3(\mathrm{kr})^2 + 9\right]^2}{(\mathrm{kr})^4\left[(\mathrm{kr})^4 + 9(\mathrm{kr})^2 + 45\right]} \tag{39}$$

ph(rad) vs kr

**Fig. 19**

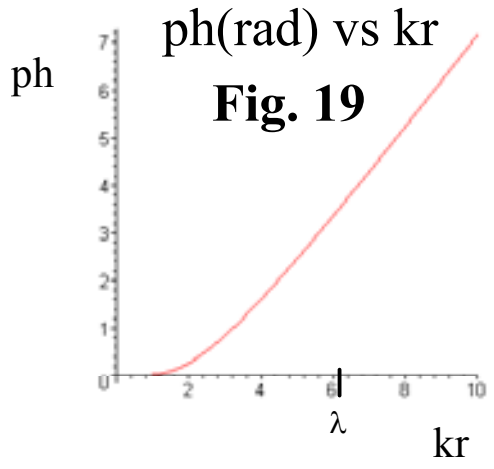

Cph/$c_o$ vs kr

**Fig. 20**

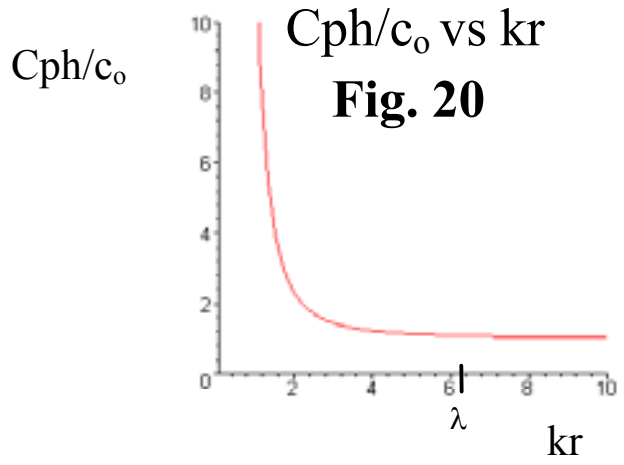

Cg/$c_o$ vs kr

**Fig. 21**

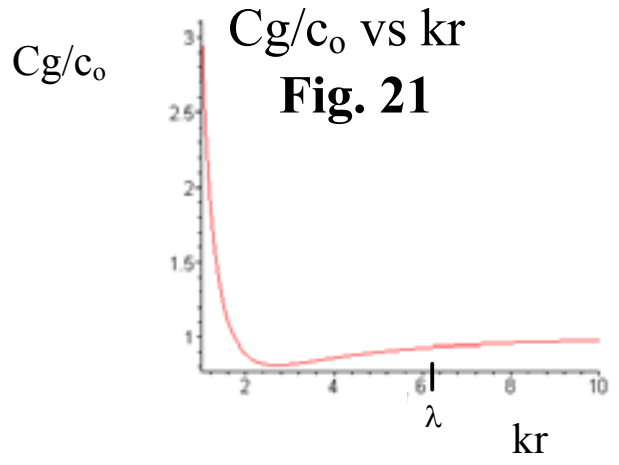

## $E_\theta$ phase, phase speed, group speed analysis

$$ph = kr + ArcTan\left[\frac{\frac{-6}{(kr)^3} + \frac{1}{kr}}{\frac{6}{(kr)^4} - \frac{3}{(kr)^2}}\right] \underset{kr \ll 1}{\approx} -\frac{1}{30}(kr)^5 + O(kr)^7 \tag{40}$$

$$Cph = c_o\left(\frac{36 - 3(kr)^4 + (kr)^6}{(kr)^6 - 6(kr)^4}\right) \tag{41}$$

$$Cg = \frac{c_o\left[36 - 3(kr)^4 + (kr)^6\right]^2}{(kr)^{12} - 3(kr)^{10} + 18(kr)^8 + 252(kr)^6 - 1080(kr)^4} \tag{42}$$

ph(rad) vs kr

**Fig. 22**

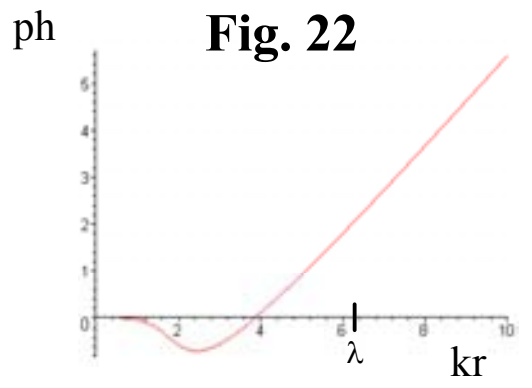

Cph/$c_o$ vs kr

**Fig. 23**

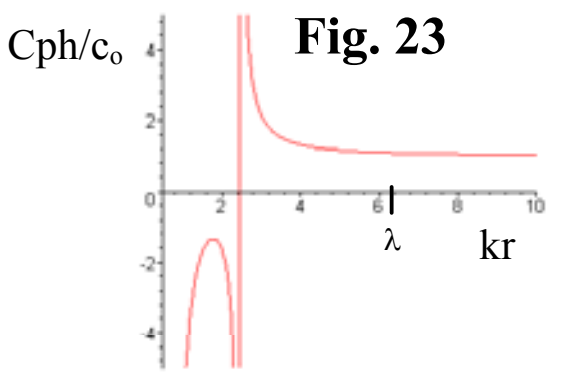

Cg/$c_o$ vs kr

**Fig. 24**

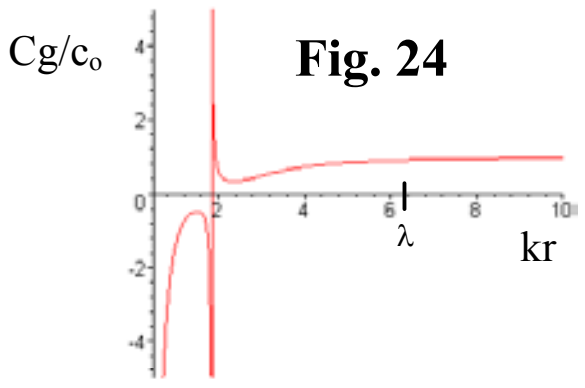

**Field contour plots** (linear quadrapole in center and vertical)

Contour plots of the fields (Eq. 31 - 33) using Mathematica software yields:

$B_\phi$ contour plot

**Fig. 25**

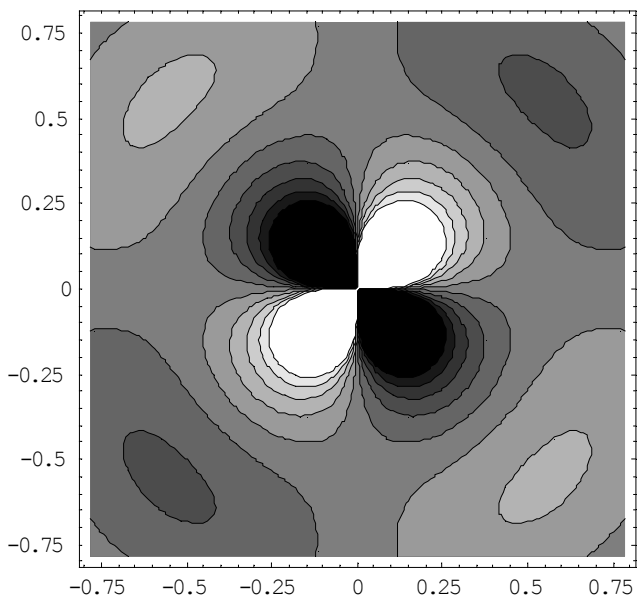

Er contour plot

**Fig. 26**

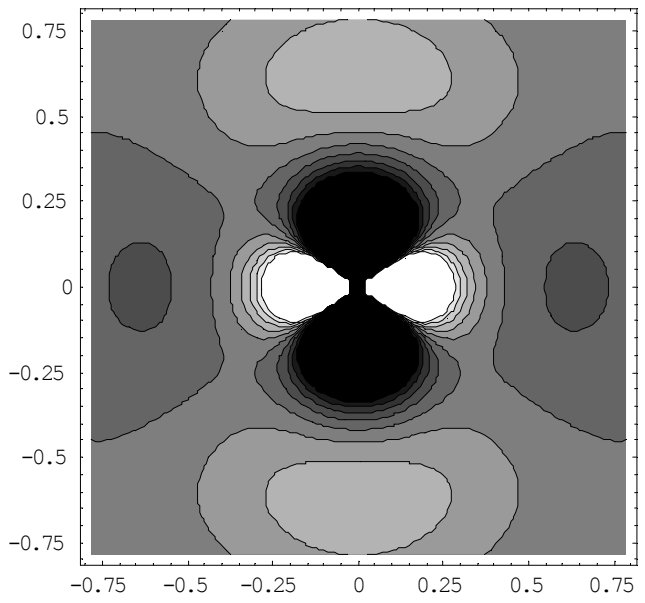

$E_\theta$ contour plot

**Fig. 27**

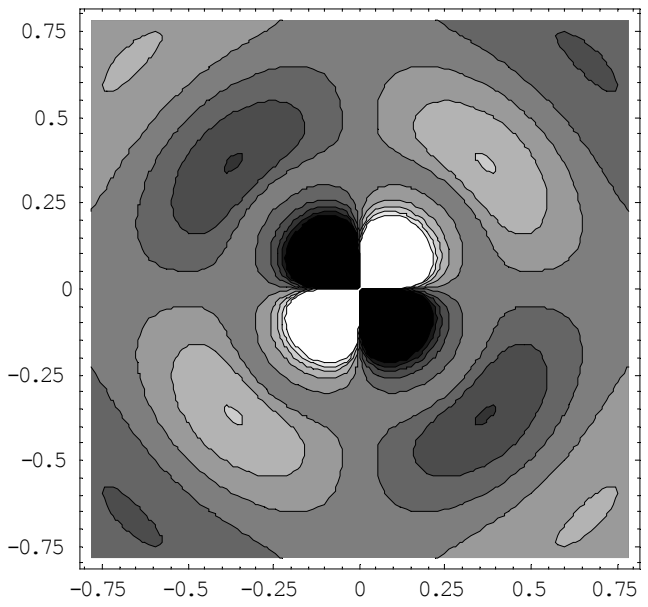

**Total E field plots** (linear quadrapole in center and vertical unless specified)

Using vector field plot graphics in Mathematica software, the Er and $E_\theta$ can be combined and plotted as vectors (Fig. 28). A more detailed plot of the total E field can be obtained by using the fact that a line element crossed with the electric field = 0. A contour plot of the resulting relation yields the total E field plots below (Fig. 29-30).

Near-field
Vector plot
**Fig. 28**

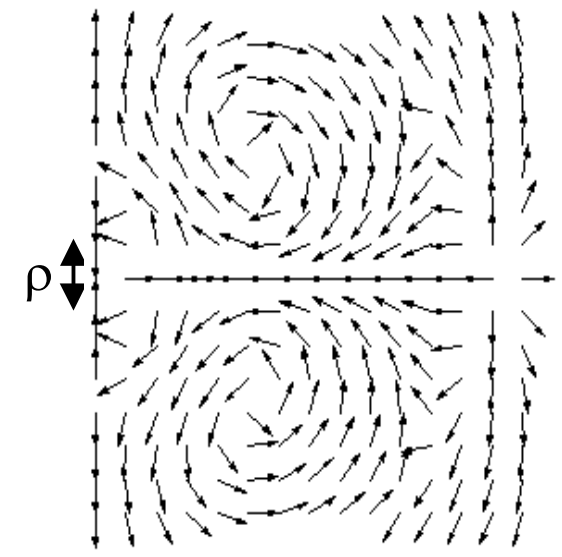

Near-field
Contour plot
**Fig. 29**

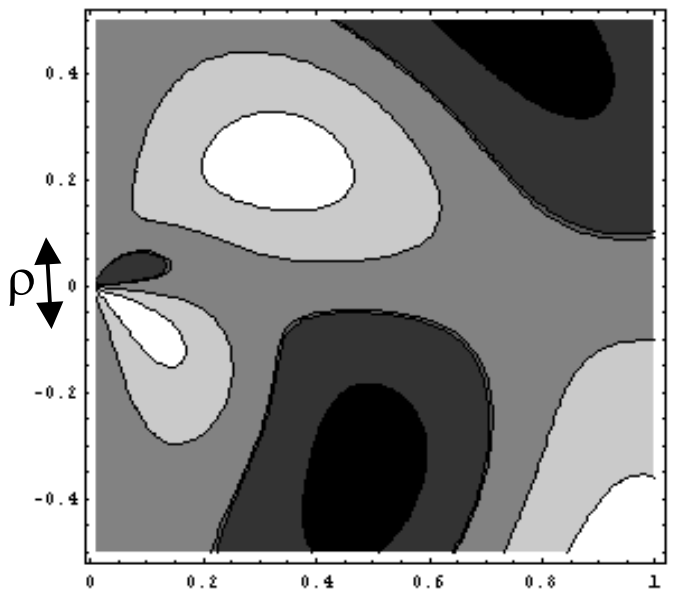

Far-field
Contour plot
**Fig. 30**

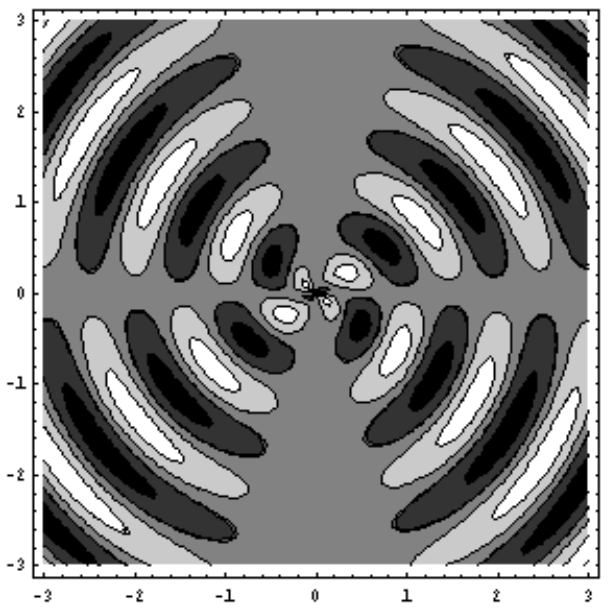

**Information speed**

If an amplitude-modulated signal propagates a distance (d) in time (t), then the information contained in the modulation propagates at a speed:

$$c_{inf} = d/(t+T) \qquad (43)$$

where (T) is the amount of time the modulated signal must pass by the detector in order for the information to be determined. The information in the wave is determined by measuring the amplitude, frequency and phase of the wave modulation envelope.

If a wave is propagated across distances in the farfield of the source, then the wave information speed is approximately the same as the wave group speed. This is because the wave propagation time (t) is much greater than the wave information scanning time (T), consequently: $c_{inf} = d/t = c_g$ .

In the nearfield of the source, if nothing is known about the type of modulation, then the scanning time (T) can be much larger than the wave propagation time (t), thereby making the wave information speed much less than the wave group speed. This can be understood by noting that several modulation cycles are required for a Fourier analyzer to be able to determine the wave modulation amplitude, frequency, and phase. But if the type of modulation is known, then only a few points of the modulated signal need to be sampled by a detector in order to curve fit the signal and therefore determine the modulation information. If the noise in the signal is very small then the signal scanning time (T) can be made much smaller than the signal propagation time (t), consequently: $c_{inf} \sim d/t = c_g$ .

**Relativistic consequences**

According to the relativistic Lorentz time transform (Eq. 44), if an information signal can propagate at a speed (w) faster than the speed of light (c), then the signal can be reflected by a moving frame (v) located a distance (L) away and the signal will arrive before the signal was transmitted ($\Delta t' < 0$). Since the information in the signal can be used to prevent the signal from being transmitted, this results in a logical contradiction (violation of causality). How can the signal be detected if it was never transmitted? Consequently, Einstein in 1907 stated that superluminal signal velocities are incompatible with Relativity theory [16].

**Fig. 31**

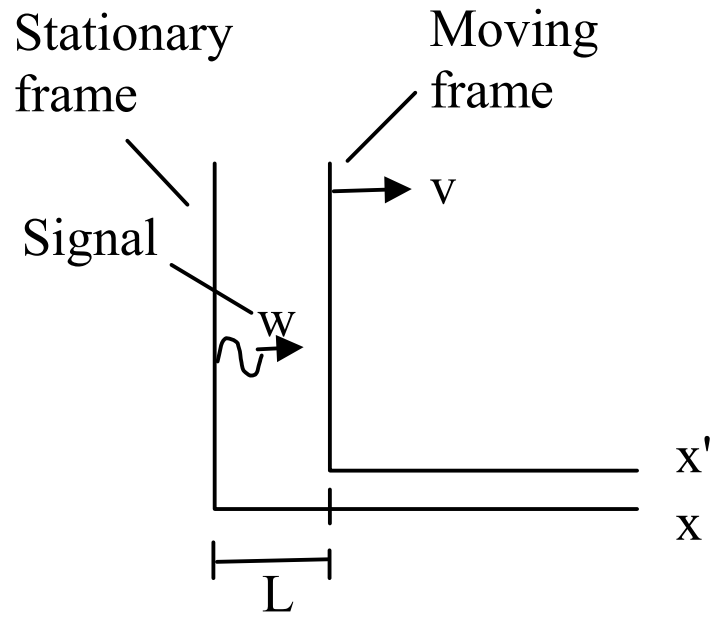

$$\Delta t' = \gamma\left(\Delta t - \frac{v}{c^2}\Delta x\right) \underset{w > \frac{c^2}{v}}{\Rightarrow} \text{Neg} \qquad (44)$$

where: $\Delta t = \dfrac{L}{w}$ $\quad \Delta x = L$

$$\gamma = \frac{1}{\sqrt{1 - \frac{v^2}{c^2}}}$$

Because Relativity theory predicts that a moving reflector which has mass can never move faster than light ($v < c$), then in order for ($\Delta t' > 0$) the signal propagation speed must be less than the speed of light.

## Conclusion

The analysis presented in this paper has shown that the fields generated by an electric or magnetic dipole or quadrapole, and also the gravitational fields generated by a quadrapole mass source, propagate superluminally in the nearfield of the source and reduce to the speed of light as they propagate into the farfield. The group speed of the waves produced by these systems has also been shown to be superluminal in the nearfield. Although information speed can be less than group speed in the nearfield, it has been shown that if the method of modulation is known and provided the noise of the signal is small enough, the information can be extracted in a time period much smaller than the wave propagation time. This would therefore result in information speeds only slightly less than the group speed which has been shown to be superluminal in the nearfield of the source. It has also been shown that Relativity theory predicts that if an information signal can be propagated superluminally, then it can be reflected by a moving frame and arrive at the source before the information was transmitted, thereby enabling causality to be violated.

Given these results, it is at present unclear how to resolve this dilemma. Relativity theory could be incorrect, or perhaps it is correct and information can be sent backwards in time. Perhaps as suggested by the 'Hawking chronology protection conjecture' [17], nature will intervene in any attempt to use the information to change the past. Therefore information can be propagated backwards in time but it cannot be used to change the past, thereby preserving causality. Another possibility is that according to the 'many-worlds' interpretation of quantum mechanics [18], multiple universes are created any time an event with several possible outcomes takes place. If this interpretation is correct, then information can be transmitted into the past of alternative universes, thereby preserving the past of the universe from which the signal was transmitted.

In addition to the theoretical implications of the research discussed above, it may also have practical applications, such as increasing the speed of electronic systems which will soon be limited by light speed-time delays. It should also be possible to reduce the time delays inherent in current astronomical observations by monitoring lower frequency EM and eventually gravitational radiation from these sources. Lastly, using low frequency EM transmissions, it should be possible to reduce the long communication time delays to spacecraft, which will become more essential as we explore our solar system and beyond.